\documentclass[aps,prd,twocolumn,showpacs,amsmath,amssymb,nofootinbib]{revtex4}
\usepackage{graphicx}
\usepackage{dcolumn}
\usepackage{bm}
\usepackage{amsmath}    
\usepackage{amsfonts}
\usepackage{amssymb}
\def\be{\begin{equation}}
\def\ee{\end{equation}}
\def\bea{\begin{eqnarray}}
\def\eea{\end{eqnarray}}

\def\hp{n}
\def\fnu{F_{\nu}}
\def\of{\overline{F}}
\def\dof{\frac{\partial{\overline{F}}}{\partial q}}
\def\pE{\frac{p}{E}}
\def\chiij{\chi_{ij} n^i n^j}
\def\chimn{\chi_{mn} n^m n^n}

\def\12{\frac{1}{2}}
\def\32{\frac{3}{2}}
\def\23{\frac{2}{3}}

\def\chiijp{\chi'_{ij} n^i n^j}
\def\dofe{\frac{\partial{\overline{F_{\nu}}}}{\partial E}}
\def\dfxi{\frac{\partial{F_{\nu}}}{\partial x^i}}
\def\dfe{\frac{\partial{F_{\nu}}}{\partial E}}

\def\fI{F^{(1)}_{\nu}}
\def\fII{F^{(2)}_{\nu}}
\def\fiI{\Phi^{(1)}}

\def\psiI{\Psi^{(1)}}
\def\psiIp{\Phi^{(1) '}}
\def\fiII{\Phi^{(2)}}

\def\psiIIp{\Psi^{(2) '}}

\def\fIk{F^{(1)}_{\bf k}}
\def\fIk1{F^{(1)}_{{\bf k}_1}}
\def\fIIk{F^{(2)}_{\bf k}}
\def\fIIpk{F^{(2) '}_{\bf k}}
\def\fiIk{\Phi^{(1)}_{\bf k}}

\def\psiIk{\Psi^{(1)}_{\bf k}}

\def\fiIIk{\Phi^{(2)}_{\bf k}}

\def\fiIk1p{\Phi^{(1) \, '}_{{\bf k}_1}}
\def\fiIk1{\Phi^{(1)}_{{\bf k}_1}}

\def\psiIk1p{\Psi^{(1) \, '}_{{\bf k}_1}}
\def\psiIk1{\Psi^{(1)}_{{\bf k}_1}}

\def\psik1{\Psi_{{\bf k}_1}}
\def\fik1{\Phi_{{\bf k}_1}}
\def\psik2{\Psi_{{\bf k}_2}}
\def\fik2{\Phi_{{\bf k}_2}}

\def\psik{\Psi_{k_1}}
\def\fik{\Phi_{k_1}}
\def\psikd{\Psi_{k_2}}
\def\fikd{\Phi_{k_2}}

\begin{document}

\title{The impact of cosmic neutrinos on the gravitational-wave background}

\author{Anna Mangilli}
\email{mangilli@ieec.uab.es}
\affiliation{Institute of Space Sciences (CSIC-IEEC)
Campus UAB, Torre C5 parell 2.
Bellaterra (Barcelona), Spain and\\
Dipartimento di Fisica `Galileo Galilei', Universit\`a di 
Padova, via Marzolo 8, I-35131 Padova, Italy}

\author{Nicola Bartolo}
\email{bartolo@pd.infn.it}
\affiliation{Dipartimento di Fisica `Galileo Galilei', Universit\`a di 
Padova and \\ 
INFN, Sezione di Padova, via Marzolo 8, I-35131 Padova, Italy}

\author{Sabino Matarrese}
\email{sabino.matarrese@pd.infn.it}
\affiliation{Dipartimento di Fisica `Galileo Galilei', Universit\`a di 
Padova and \\ 
INFN, Sezione di Padova, via Marzolo 8, I-35131 Padova, Italy}

\author{Antonio Riotto}
\email{antonio.riotto@pd.infn.it}
\affiliation{INFN, Sezione di Padova, via Marzolo 8, I-35131 Padova, Italy and 
\\ CERN, Theory Division, CH-1211 Geneva 23, Switzerland}

\date{\today}

\begin{abstract}
We obtain the equation governing the evolution of the
cosmological gravitational-wave background, accounting for the 
presence of cosmic neutrinos, up to second order in perturbation theory. 
In particular, we focus on the epoch during radiation dominance,   
after neutrino decoupling, when 
neutrinos yield a relevant contribution to the total energy density and 
behave as collisionless ultra-relativistic particles.  
Besides recovering the standard damping effect due to neutrinos, 
a new source term for gravitational waves is shown to arise from 
the neutrino anisotropic stress tensor. 
The importance of such a source term, so far completely disregarded 
in the literature, is related to the high velocity dispersion of neutrinos 
in the considered epoch; its computation requires solving the full 
second-order Boltzmann equation for collisionless neutrinos. 

\end{abstract}

\maketitle

\section{Introduction}
An important discriminator among different models for the generation of 
the primordial density perturbations is the 
level of the gravitational-wave background predicted by these models. 
For example, within the inflationary scenario 
the tensor (gravitational-wave) amplitude generated by tiny initial 
quantum fluctuations during the accelerated inflationary expansion of 
the universe depends on the energy scale at which this inflationary 
period took place, and it can widely vary among different inflationary 
models~\cite{review1,review2}. On the other hand, some alternative 
scenarios, such as the curvaton model, typically predict 
an amplitude of primordial tensor modes that is far too small to be ever 
detectable by future satellite experiments aimed at observing the 
B-mode of the Cosmic Microwave Background polarization. 

There is however another background of stochastic gravitational waves 
of cosmological origin. Gravitational waves (as well as vector modes) 
are inevitably generated at second order in perturbation theory by scalar 
density perturbations~\cite{secgw1,secgw3,secgw4,secgw5,secgw6}. 
This is due to the 
fact that the non-linear evolution always involves quadratic 
source terms for tensor (and vector) perturbation modes made of 
linear scalar (density) perturbations. 

Since the level of density perturbations is well determined by CMB 
anisotropy measurements and 
large-scale structure observation~\cite{WMAP3,WMAP5}, we know that these
secondary vector and tensor modes (produced after the primordial curvature 
perturbations have been generated) must exist and their amplitude must have a 
one-to-one relation with the level of density perturbations. In this 
sense, the scalar-induced contribution can be computed directly from 
the observed density perturbations and general relativity, and is 
independent of the specific cosmological model for generating the 
perturbations.\footnote{See however Ref.~\cite{curvatontensor}, where in the context of the curvaton mechanism, 
second-order gravitational waves can be produced when the perturbations are still of isocurvature nature, 
thus resulting to be strongly model dependent.}  

Such a background of gravitational waves could be interesting in relation 
to future high-sensitivity CMB polarization experiments or for 
small-scale direct detectors, such as the space-based laser interferometer 
Big Bang Observer (BBO) and the Deci-hertz Interferometer 
Gravitational Wave Observatory (DECIGO) operating in the frequency 
range $0.1$ -- $1$Hz~\cite{BBODECIGO} with an improved  
sensitivity (in terms of the closure energy density of gravitational waves,  
$\Omega_{GW}\sim 10^{-17}-10^{-15}$). In particular, in 
Ref.~\cite{MHM} the effects of secondary tensor and vector modes on
the large scale CMB polarization have been computed, showing that they 
dominate over the primordial gravity-wave background if 
the tensor-to-scalar perturbation ratio on large scales is $r <  10^{-6}$. 
More recently, Ref.~\cite{Anandaetal} computed the power-spectrum of the 
secondary tensors accounting for their evolution during the 
radiation dominated epoch, to see their effects on the scales relevant 
for small-scale direct detectors, and Ref.~\cite{BST} extended this 
analysis by accounting for a more detailed study of the transfer function 
for the secondary tensor modes. 

In this paper we consistently account for the presence of cosmic 
neutrinos to analyze their impact on the evolution of the second-order 
gravitational-wave background. 
At linear order it has been shown that there is a 
damping effect due to the anisotropic stress of free-streaming neutrinos 
that strongly affects the primordial gravitational-wave background on 
those wavelengths which enter the horizon during the radiation 
dominated epoch (at the level of 
$30\%$)~\cite{Bond,RS,D,Weinb,Kamion,Repko,Komatsu} (see also Ref.~\cite{kasaitomita}). 
At second order, along with the analogous damping effect, we find 
that free-streaming neutrinos are an important source for the 
second-order gravitational-wave background during the radiation-dominated 
epoch. We find completely new source terms,  
arising because of the fact that neutrinos give a relevant 
contribution to the total energy density during this epoch 
and they behave as ultra-relativistic collisionless particles 
after their decoupling: their high velocity dispersion acts as an 
extra source for the 
second-order gravitational waves. To compute such a contribution we 
evaluate the 
second-order tensor part of the neutrinos' anisotropic stress tensor, 
that has been neglected so far. This is achieved by computing and 
solving the Boltzmann equation for neutrinos. 
Approximating the neutrino contribution as a perfect fluid of 
relativistic particles during the radiation era leads to a 
serious underestimate of their role. 
Let us stress that the new contribution is at 
least of the same order of magnitude as that computed by 
adopting a fluid treatment in the source of the scalar-induced 
gravitational waves. Moreover it has a clear physical interpretation. 
It arises in the Boltzmann equation  
from a ``lensing'' effect of the neutrinos as they travel through the 
inhomogeneities of the gravitational 
potential. 

The paper is organized as follows. In Secs.~\ref{GP} and ~\ref{IIGWeq} 
we derive the evolution equation for the tensor (gravitational-wave modes) 
at second order, accounting for photons and neutrinos. 
In Sec.~\ref{Bn} we present 
the Boltzmann equation for neutrinos approximated as being 
collisionless massless  particles 
(see Appendix A, B and C for details about the Boltzmann equation for massive 
neutrinos) and give an integral solution for it. 
Sec.~\ref{main} contains the computation needed 
to determine the tensor part of the second-order anisotropic stress tensor 
of neutrinos, which leads to one of our main results, Eq.~(\ref{new}). 
Finally, in Sec.~\ref{photons+neutrinos} we derive the photon contribution, 
consistently accounting for the presence of neutrinos, which 
leads also to a new source of gravitational waves. 
In Sec.~\ref{Concl} we present the summary and our main conclusions.     

%%%%%%%%%%%%%%%%%%%%%%%%%%%%%%%%%%%%%%%%%%%%%%%%%%%%%%%%%%%%%%%%%%%%%%%%%%%%%
\section{Second-order Gravitational Waves}
%%%%%%%%%%%%%%%%%%%%%%%%%%%%%%%%%%%%%%%%%%%%%%%%%%%%%%%%%%%%%%%%%%%%%%%%%%%%

%%%%%%%%%%%%%%%%%%%%%%%%%%%%%%%%%%%%%%%%%%%%%%%%%%%%%%%%%%%%%%%%%%%%%%
\subsection{Metric perturbations in the Poisson gauge}
\label{GP}
%%%%%%%%%%%%%%%%%%%%%%%%%%%%%%%%%%%%%%%%%%%%%%%%%%%%%%%%%%%%%%%%%%%%%%%%%%

The second-order metric perturbations around a flat
Friedmann-Robertson-Walker (FRW) background can be described by 
the line-element in the Poisson gauge
\bea \label{poisson}
ds^2&=&a^2(\tau)\left[-e^{2\Phi} d\tau^2+2\omega_i dx^i d\tau +
  (e^{-2\Psi}\delta_{ij}
\right. \nonumber \\
&+&\left. \chi_{ij}) dx^i dx^j\right]\, .
\eea
In this gauge one scalar degree of freedom is eliminated from $g_{0i}$ and 
one scalar and two vector degrees of freedom are removed from $g_{ij}$.
As usual $a(\tau)$ is the scale factor and
$\tau$ is the conformal time.  
The functions $\Phi$ and $\Psi$ 
are scalar functions which
correspond to the Newtonian potential and to the spatial curvature
perturbations, respectively. Within second-order perturbation theory, they 
consist in the sum of a linear and a second-order term, such that $\Phi$ and
$\Psi$ can be written as 
\be\label{II-fipsi}
\Phi=\Phi^{(1)}+\Phi^{(2)}/2 \quad \textrm{and} 
\quad \Psi=\Psi^{(1)}+\Psi^{(2)}/2\, .
\ee
Since the choice of the exponentials greatly helps in simplifying the computation of many expressions, they will be kept where it is convenient. It is worth remarking here that all the equations in the following where the exponential show up are meant to be second order equations, therefore the exponentials are to be thought as implicitly truncated up to second order in all these expressions e.g. $e^{2\Phi} \simeq 1 + 2 \Phi^{(1)} + \Phi^{(2)} + 2 (\Phi^{(1)})^2$.

The remaining functions that appear in Eq.~(\ref{poisson}) account for
second-order vector ($\omega_{i}$) and tensor ($\chi_{ij}$) modes. 
Tensor perturbations are traceless and transverse: $\chi^i_i=0$,
$\partial_{i}\chi^{ij}=0$ and vectors have vanishing spatial
divergence: $\partial^i \omega_i = 0$. 
Linear vector modes have been neglected as they are not produced by 
standard mechanisms, such as inflation, that generate 
cosmological perturbations \cite{MatBar},
\cite{Matinfl}. As discussed, for example in Refs.~\cite{MatBar} 
and~\cite{Matinfl}, indeed linear vector modes have
decreasing amplitudes and they are not generated in the presence of scalar
fields, while the first-order tensor part gives a negligible
contribution to second-order perturbations. 
Second-order vector and tensor modes however must be
taken into account, even if they were initially zero. 
This is because scalar, vector and tensor modes
are dynamically coupled at this stage and second-order vectors and tensors are
generated by first-order scalar mode-mode coupling. First-order
perturbations behave as a source for the intrinsically second-order
fluctuations \cite{secgw4}.

Since our main task is to provide the second-order Einstein's 
equations that describe the evolution of tensor modes, we are
interested in the spatial components of both the Einstein and
the energy-momentum tensors.
Here we find that, using the Christoffel symbols
obtained in Appendix~\ref{christoffel} and accounting only for the terms up to second order, the spatial Einstein tensor reads
\bea\label{2oein}
G^i_{~j}&=&\frac{1}{a^2}\biggl[e^{-2\Phi}\biggl({\mathcal H}^2-2\frac{a''}{a}-
2\Psi'\Phi'-3(\Psi')^2+\\
&+& 2{\mathcal H}\bigl(\Phi'+2\Psi'\bigr)+2\Psi''\biggr) 
+ e^{2\Psi}\biggl(\partial_k\Phi\partial^k\Phi+\nonumber\\
&+& \nabla^2\Phi-\nabla^2\Psi
\biggr)\biggr]\delta^i_j 
+ \frac{e^{2\Psi}}{a^2} 
\biggl(-\partial^i\Phi\partial_j\Phi  -
\partial^i\partial_j\Phi+\nonumber\\
&+&\partial^i\partial_j\Psi- 
\partial^i\Phi\partial_j\Psi
+\partial^i\Psi\partial_j\Psi-\partial^i\Psi\partial_j\Phi\biggr)\nonumber\\
&-&\frac{{\mathcal H}}{a^2}\left(\partial^i\omega_j+\partial_j\omega^i\right)
-\frac{1}{2a^2}\left(\partial^{i}\omega_j'+\partial_j\omega^{i'}\right)
\nonumber \\
&+& \frac{1}{a^2}\left(
{\mathcal H}\chi^{i'}_j+\frac{1}{2}\chi_j^{i''}-\frac{1}{2}\nabla^2
\chi^i_j\right) \, ,\nonumber
\eea
where $ \mathcal H =a'/a$, and a prime denotes differentiation 
w.r.t. conformal time.

%%%%%%%%%%%%%%%%%%%%%%%%%%%%%%%%%%%%%%%%%%%%%%%%%%%%%%%%%%%%%%%%%%%%%%%%%
\subsection{Second-order gravitational-wave evolution 
equation during the radiation-dominated era}\label{IIGWeq}
%%%%%%%%%%%%%%%%%%%%%%%%%%%%%%%%%%%%%%%%%%%%%%%%%%%%%%%%%%%%%%%%%%%%%%%%%

In Fourier space the equation which describes the evolution of 
second-order gravitational waves (GW) can be 
put in the form:
\be\label{eqevol}
 \chi^{''}_{k,\lambda}+2{\cal H} \chi^{'}_{k,\lambda}
+k^{2}\chi_{k,\lambda}=16\pi G a^{2} S_{k,\lambda}\, ,
\ee
where the subscript $\lambda$ refers to the two possible polarization
states of a gravitational wave.
Each mode $\chi_{k}$ is in fact transverse with respect to the direction 
along which it propagates and, for a mode traveling in the z direction, 
$\chi_{ij}$ can be written as:
\begin{equation}
 \chi_{ij}=
 \left( \begin{array}{ccc}
 \chi_{+} & \chi_{\times} & 0 \\
 \chi_{\times} & -\chi_{+} & 0 \\
  0 & 0 & 0
 \end{array} \right)\, .
 \end{equation}
The two degrees of freedom account for the two polarization states
$\chi_{+}$ and $\chi_{\times}$.  

The source term $S_{k,\lambda} $ for GW in the radiation era
consists in the sum of three different parts: $E_{k,\lambda}$ that 
comes from the Einstein tensor, $\Pi^{(\nu)}_{k,\lambda}$ that comes from 
the neutrino anisotropic stress tensor term and $\Pi^{(\gamma)}_{k,\lambda}$ 
that accounts for the photon contribution. We then have:
\be
S_{k,\lambda}= E_{k,\lambda} + \Pi^{(\nu)}_{k,\lambda} + 
\Pi^{(\gamma)}_{k,\lambda} \;.
\ee

In making the source term $S_{k,\lambda}$ explicit, the first step
is to extract the tensor part of the Einstein and energy-momentum tensors 
to get the corresponding transverse traceless component.

This can be done by making use of the
projection operator ${\cal P}_{rj}^{~is}$, so that \footnote{
In the following, where not necessary, we will omit the $TT$ subscript.}:
\be
(\Pi^i_{~j})_{TT}={\cal P}_{rj}^{~is} T^r_{~s}\, .
\ee

Here $T^r_{~s}$ contains both the neutrino and photon
contributions
 \be
T^r_{~s}= T^{r(\nu)}_{~s}+ T^{r(\gamma)}_{~s} \; .
\ee
The definition of such an operator is given in Ref.~\cite{secgw5}
\begin{equation} 
{\cal P}_{rj}^{~is}={\cal P}^i_r {\cal P}^s_j-\frac{1}{2}
{\cal P}^i_j {\cal P}^s_r\, ,
\end{equation}
where 
\begin{equation}
{\cal P}^i_j =\delta^i_j-\left( \nabla^2 \right)^{-1} \partial^i\partial_j\, .
\end{equation}
Moving to Fourier space the two-indices operator ${\cal P}^i_j$ reads
\be
 \widetilde{\cal P}^i_j =( \delta^i_j - \hat{k}^i \hat{k}_j)\, ,
\ee
and then the operator to apply is 
\bea\label{tensor-operator}
\widetilde{\cal P}_{rj}^{~is} &=&
 \delta^i_r \delta^s_j - \delta^i_r \hat{k}^s \hat{k}_j 
 - \delta^s_j \hat{k}^i \hat{k}_r + \12 \hat{k}^i \hat{k}_r \hat{k}^s
 \hat{k}_j \nonumber\\
&-& \12 \delta^i_j \delta^s_r +\12
\delta^i_j \hat{k}^s \hat{k}_r + \12 \delta^s_r \hat{k}^i  \hat{k}_j.
\eea
 
The contribution to the source term coming from the Einstein tensor
can then be written as:

\bea
E^i_{~j} &=& -\frac{1}{a^2} {\cal P}_{rj}^{~is} 
\biggl(-\partial^r\Phi\partial_s\Phi  -
(\partial^r\partial_s\Phi) 2 \Psi 
+2 (\partial^r\partial_s\Psi) \Psi \nonumber \\
&-& \partial^r\Phi\partial_s\Psi
+\partial^r\Psi\partial_s\Psi-\partial^r\Psi\partial_s\Phi\biggr)\, .
\eea

To get $E_{k,\, \lambda}$ we now have to move to Fourier space and 
project along one of the two polarization states
\mbox{$\lambda= + \, , \times$}.

If we take an orthonormal basis made up by the three unit 
vectors $\bf{e}$, $\bf{\bar{e}}$ and $\bf{\hat{k}}$, 
the two polarization tensors are defined as follows:
\bea\label{tensorpolarizz}
\varepsilon^+_{ij}(\textbf{k})= \frac{1}{\sqrt2}[e_i(\textbf{k}) 
e_j(\textbf{k}) -
  \bar{e}_i(\textbf{k}) \bar{e}_j (\textbf{k})]\, ,\\
\varepsilon^\times_{ij}(\textbf{k})= \frac{1}{\sqrt2}[e_i(\textbf{k})
  \bar{e}_j(\textbf{k})
 + \bar{e}_i(\textbf{k}) e_j(\textbf{k})]\nonumber \, .
\eea

Remembering that a gravitational wave is
transverse with respect to the direction $\hat{\bf k}$ along 
which it propagates we find that 
\footnote{
The indices of the 3-D unit vectors are 
lowered and raised by using $\delta_{ij}$ so that 
\mbox{$\delta^i_j e^i e_j= 1 = \delta^i_j \bar{e}^i \bar{e}_j$}.}
:
\bea
E_k^\lambda &=& \varepsilon^{j \, \lambda}_{i}E^i_{~j} =
 a^{-2} \int \frac{d^3 k_1 \, d^3 k_2}{(2 \pi)^3} \,\delta({\bf k}_1 + 
{\bf k}_2-
 {\bf k}) \nonumber\\
&\times&  \varepsilon^{j \, \lambda}_{i} k_{1 \, j} k^i_2 \Big[k_1 k_2 \fik \fikd - 
2 k_1^2 \fik \psikd + 2 k_1^2 \psik \psikd \nonumber\\
&+& k_1 k_2 \fik \psikd - k_1 k_2 \psik \psikd + k_1 k_2 \psik \fikd \Big]\, .
\eea
The explicit form for the product $ \varepsilon^{j \, +}_{i} k_{1 \, j} k^i_2$ is given by the analogous of eq. (\ref{enn}).

The next sections are devoted to the computation of the neutrino source term 
$\Pi^{(\nu)}_{k,\lambda}$, since the term due to the photons is already 
known and we will recall 
it later. Notice however that Sec.~\ref{photons+neutrinos} 
presents a non trivial case accounting for both neutrinos and photons: here in fact the contribution of the first-order collisionless neutrinos is fully accounted for the first time when explicitly calculating the second-order photon quadrupole in the tight coupled limit (Eq.~\ref{photonquadrupole}).

%where in particular  
%the neutrinos contribution at second-order is fully accounted for the first time.   

We will proceed by steps. 
First we solve the second-order Boltzmann equation for the neutrinos in order to give the expression for the neutrino energy 
momentum tensor. Then we will  give the expression for the spatial components of the energy 
momentum tensor of neutrinos in terms of the their distribution function perturbed up to second order, and finally we can extract the transverse and tracelees 
part of it.

%%%%%%%%%%%%%%%%%%%%%%%%%%%%%%%%%%%%%%%%%%%%%%%%%%%%%%%%%%%%%%%%%%%%%%%%%%%%%
\section{Neutrinos and second-order tensor modes}
%%%%%%%%%%%%%%%%%%%%%%%%%%%%%%%%%%%%%%%%%%%%%%%%%%%%%%%%%%%%%%%%%%%%%%%%

\subsection{Solution of the second-order Boltzmann equation for neutrinos}
\label{Bn}

The Boltzmann equation up to second-order for decoupled neutrinos can be
written as

\be \label{bolt-II}
 \frac{\partial{F_{\nu}}}{\partial{\tau}}
+\frac{\partial{F_{\nu}}}{\partial{x^{i}}}\cdot\frac{dx^{i}}{d \tau}
+\frac{\partial{F_{\nu}}}{\partial{E}}\cdot\frac{dE}{d \tau} 
+\frac{\partial{F_{\nu}}}{\partial{n^i}}\cdot\frac{d n^i}{d\tau}
=0 \, ,
\ee
where $\tau$ is the conformal time, $E = \sqrt{m^2_{\nu} + p^2}$, $p^2=g_{ij} P^i
P^j$ is the squared neutrinos 3-momentum and $P^{\mu}$ is the neutrino 4-momentum defined as $P^{\mu}= d
x^{\mu} / d \lambda$. Here $\lambda$ parametrizes the particle's path and $x^{\mu}=(t, \bold{x})$ represents a space-time point. The unit vector $\bold{n}$, with components $n^i$, represents the neutrino momentum direction and it is therefore such that: $\bold{p}=p \bold{n}$ and $n^in^j\delta_{ij} = 1$.
Equation (\ref{bolt-II}) refers to the general case of massive neutrinos ($g_{\mu
  \alpha} P^{\mu} P^{\alpha}= - m^2_{\nu}$).

The neutrino distribution function $F_{\nu}$ at this stage will
consist in an unperturbed, a first-order and a second-order term, 
such that we can put it in the form
\be\label{II-distrib}
F_{\nu}=\overline{F}_{\nu}(E, \tau) + F^{(1)}_{\nu}(E, \tau, x^i,n^i)
 + \frac{F^{(2)}_{\nu}}{2}(E, \tau, x^i, n^i).
\ee
% The components of the 4-momentum $P^{\mu}$ can be expressed in terms
% of the perturbed scalars, vectors and tensors so that, with the metric
% (\ref{poisson}), we have:
% \bea \label{4mom}
% P^i&=& \frac{p \hp^i}{a} [1+ \Psi - \12 \chiij] \quad i,j= 1,2,3\nonumber\\
% P^0&=& \frac{E}{a} [1- \Phi + \omega_i \pE \hp^i].
%  \eea
% The details about the Boltzmann equation up to second order in the 
% perturbations for massive neutrinos are contained 

The explicit forms for the different contributions in
Eq.~(\ref{bolt-II}) are listed in Appendix~\ref{appendixBoltzmannII}.

Since we are interested in evaluating the neutrino contribution during
the radiation-dominated epoch, we can specialize 
the second-order Boltzmann equation for ultrarelativistic (massless) neutrinos.
Replacing $E$ with the comoving 3-momentum $q$ as one of the independent
variables in Eq.~(\ref{bolt-II}) such that
$p=E \quad \textrm{and} \quad q=a p$, the relevant 
equation takes the form:
\bea \label{bolt-II-rel}
\12 \frac{\partial \fII}{\partial \tau} &+& \12 \frac{\partial
\fII}{\partial x^i} \hp^i = - \frac{\partial \fI}{\partial x^i} \hp^i
(\fiI + \psiI) \nonumber\\
&-& q \frac{\partial \fI}{\partial q}(- \hp^j \fiI_{,j} + 
\Psi^{(1) '}) \nonumber\\
&-& \frac{\partial \fI}{\partial \hp^i} \, [\hp^i \hp^r (\fiI_{, r}+
\psiI_{, r}) - \Psi^{(1) i}_{\phantom{1} ,} - \Phi^{(1)
  i}_{\phantom{1} ,}] \nonumber\\
&-& q \frac{\partial \overline{F}}{\partial q}\Big[-\hp^j (\12
  \Phi^{(2)}_{, j} + \fiI_{, j}(\psiI + \fiI))  \nonumber\\
&+& \12 \Psi^{(2) '} + \, V_{II} \,  - \12 \chiijp \Big] .
\eea
Going to Fourier space it can be put in the form:
\be
\fIIpk + i k \mu  \fIIk = G_{\bf k}(\tau)- T_{\bf k}(\tau)\, ,
\ee
where 
\begin{equation}
\mu=\hat{{\bf k}} \cdot \hat{{\bf n}}\, ,
\end{equation}
and 
\bea \label{G}
 G_{\bf k}( \tau)&=& - q \dof \left(-i k \mu  \fiIIk +  \Psi^{(2)'}_k+2 V_{II} \right)\\
&-& 2  \int \frac{d^3 k_1 \, d^3 k_2}{ (2 \pi)^3} \, \delta^D (k_1 + k_2
-k)
\Big[ (\fiI_{k_2} + \psiI_{k_2})\nonumber\\
&\times& \Big(i k_1 \mu \fIk1 -i {\bf k}_2 \cdot {\bf n}
  \frac{\partial \fIk1}{\partial \hp^i} \hp^i  \nonumber \\
&+& q \dof i k_1 \mu \fiI_{k_1} \Big)
+ q \frac{\partial \fIk1}{\partial q} (- i k_2 
\mu \fiI_{k_2} + \psiI_{k_2})\Big]\nonumber \, .
\eea
Here $V_{II}$ accounts for the second-order vector perturbations modes, 
see Eq.~(\ref{VII}). 
We will keep it in this implicit form, since we already know that second-order vectors 
do not take part to the tensor contribution we are interested in.

The term $T_{\bf k}$ represents the
``pure'' tensor contribution and is defined by
\be\label{puretensor}
T_{\bf k} = q \dof \sum_{\lambda=\times,+} \chi'_{{\bf k},\lambda}(\tau)
\varepsilon^{\lambda}_{sr} \hp^s \hp^r.
\ee 

The Fourier expansion for the tensor modes reads
\be 
 \chi_{ij}(\boldsymbol{x},\tau)=\sum_{\lambda=\times,+}
\int{\frac{d^{3}k}{(2\pi)^{3}}\chi_{{\bf k},\lambda}(\tau)e^{i\boldsymbol{kx}}
\varepsilon^{\lambda}_{ij}}
\ee
where $\varepsilon^{\lambda}_{ij}$ are the polarization 
tensors defined in Eq.~(\ref{tensorpolarizz}) with $\lambda=\times,+$ 
accounting for the two possible polarization states of a gravitational wave.

At this point, following the standard procedure~\cite{SZ,Dodelson}, 
we can write down an integral solution of the second-order Boltzmann 
equation as
\be
\fIIk(\tau)=\int^{\tau}_{\tau_{dec}} d \tau' e^{i k \mu (\tau' -
  \tau)}[ G_k(\tau')- T_k(\tau') ].
\ee 
As usual, $\tau_{dec}$ refers to the neutrino decoupling conformal 
time \footnote{Notice that we are assuming that at a time right
before $\tau_{dec}$ the distribution of the neutrinos is the homogeneous 
one since they are still in thermal equilibrium.}.

%%%%%%%%%%%%%%%%%%%%%%%%%%%%%%%%%%%%%%%%%%

For $\fIk1$ we take the formal solution of the first order neutrino
Boltzmann equation integrated by parts:
\bea \label{fik1}
\fIk1(\tau)&=& q \dof
 \Big[
     \fiI_{k_1}(\tau) - \int^\tau_{\tau_{dec}} d
      \tau' \, e^{i k_1 \mu_1 (\tau'-\tau)} \times \nonumber\\
    &\times& 
      (\Phi^{(1) \, '}_{k_1}
      (\tau') + 
      \Psi^{(1) \, '}_{k_1}
        (\tau'))
 \Big].
\eea
where $\mu_1= \hat{{\bf k}}_1 \cdot {\bf n}$. In order not to weigh 
further the notation, $\fIk1$ will be explicitly inserted only later, 
when strictly necessary.

\subsection{The second-order neutrino energy-momentum tensor}

The contribution to the energy momentum tensor of a given
species ``$i$'' is
\be \label{en-mom-neu}
T^{\mu (i)}_{~\nu}=g_{i}\frac{1}{\sqrt{-g}}\int{\frac{d^{3}P}{(2\pi)^{3}}
\frac{P^{\mu}P_{\nu}}{P^{0}}F_{i}}.
\ee
where $P^\mu = dx^\mu/d\lambda$ 4-momentum of the particle and $F_i$ is the distribution fucntion of 
the given species.

If we now want an expression for the second-order spatial component of the
neutrino energy momentum tensor, we find that it will consist of 
the sum of four parts. There are in fact four terms 
which contain the perturbations:
the product of the 3-momentum $P^i P_j$, the component $P^0$,
the distribution function $F_{\nu}$ and the determinant of the
metric (\ref{poisson}), $g$.
As far as the first and the second term are concerned we must remember that
\be\label{4mom1}
P^i= \frac{\overline{q}^i}{a} e^\Psi \left(1- \12 \chi_{m l} n^m n^l \right)\, ,
\ee
\bea
P_j &\equiv& g_{\mu j}P^\mu \\ 
&=& q \Big[ 2 \omega_j + \Big(e^{-\Psi} n_j - \12 \chimn n_j + \chi_{i j} n^i \Big) \Big]\, ,\nonumber 
%\nonumber
\eea
\be\label{4mom2}
P^0= \frac{q}{a^2}\, e^{- \Phi}\left(1+ \omega_i n^i  \right)\, ,
\ee
where we define $q^2=a^2 g_{ij} P^iP^j$ and we introduce the momentum 
${\bf q}=q \bf{n}$ of magnitude $q$ and direction 
$n^i$, see the notations of Ref.~\cite{Bert} and~\cite{CMBI}. 
The overline refers to unperturbed quantities and we have
\mbox{$\overline{q}^i= a^{-1} \, q n^i$}, \mbox{$\overline{q}_j = 
a q n_j$} and \mbox{$\overline{q}^0= q / a$}.
 
The determinant of the metric, $g$, up to second order is such that:
\be
(- g)^{-\12}=a^{-4} e^{3 \Psi - \Phi}
\ee
while for the distribution function we use the decomposition in
Eq. (\ref{II-distrib}). 
By performing the variable change $P_j \rightarrow q_j$ in order to 
make all the perturbations explicit in eq. (\ref{en-mom-neu}) and 
by combining all the terms,  we find that the neutrino 
energy momentum tensor at second order in perturbation theory reads:
\be \label{en-momII} 
\left( \delta T^i_{~j}\right)^{(2)}_\nu = a^{-4} g_i \int \frac{d^3 q}{(2 \pi)^3}\,q \, n^i n_j \, \fII.
\ee
Similar to the linear case, vectors and tensors 
do not show up because of the angular integration.

With this results we are able to express the spatial component of 
the second-order energy-momentum tensor Eq.~(\ref{en-momII}) in Fourier space:
\bea \label{en-momIIfourier}
\left(\delta T^{i~(\nu)}_{~j}\right)^{(2)}_{\bf k} &=&  2 a^{-3} 
\int \frac{d^3 q}{(2\pi)^3}\,q \, \hp^i \hp_j \\
&\times&
\Big[ \int^{\tau}_{\tau_{dec}} d \tau' e^{i k \mu (\tau' -\tau)} 
     [G_k(\tau')- T_k(\tau')] \Big]\nonumber
\eea

%%%%%%%%%%%%%%%%%%%%%%%%%%%%%%%%%%%%%%%%%%%%%%%%%%%%%%%%%%%%%%%%%%%%%%%%%%%%%%%
\subsection{The neutrino contribution to the source term.}
\label{main}
%%%%%%%%%%%%%%%%%%%%%%%%%%%%%%%%%%%%%%%%%%%%%%%%%%%%%%%%%%%%%%%%%%%%%%%%%%%%%%%

In order to get the transverse traceless part of the neutrino 
energy-momentum tensor we now have to make use of the operator defined in Eq.~(\ref{tensor-operator}).
Before proceeding notice that we are interested only in the tensor contribution to the energy momentum tensor. 
Therefore it is 
very useful to keep in mind the decomposition of the distibution function into its scalar, vector and 
tensor parts, according to the splitting of Ref.~\cite{kasaitomita}. For example the tensor part is given by 
\begin{eqnarray}
\label{splitting}
\delta F= \sum_{\lambda} f_\lambda({\bf k},\tau,q,{\hat{\bf n}}) \epsilon^{\lambda}_{ij} n^i n^j\, . 
\end{eqnarray}
This greatly helps in simplifying all the expressions: it is telling us that 
the only tensor contributions to the energy momentum tensor 
are those that can be built out of the product of two momentum direction $n^i$. Looking back to 
Eq.~(\ref{en-momII}) and Eq.~(\ref{bolt-II-rel}) we see that, 
apart from the straightforward term depending on the gravitational waves $\chi_{ij}$, 
there is just one term of this type, namely
\begin{equation}
\label{only}
- \frac{\partial \fI}{\partial \hp^i} \, [\hp^i \hp^k (\fiI_{, k}+
\psiI_{, k})]\, .
\end{equation}
Therefore, from now on we will focus 
just on this term in Eq.~(\ref{en-momIIfourier}), and we will drop all the 
others since they correspond to scalar and vector modes.\footnote{
In fact we have explicitly verified that all the 
remaining terms vanish once the energy momentum tensor is projected along the two polarization 
tensors~(\ref{tensorpolarizz}) and the angular integration in Eq.~(\ref{en-momIIfourier}) is performed.}
It is interesting to notice that the term~(\ref{only}) has a clear and simple physical interpretation. 
It arises in the Boltzmann equation  
from a ``lensing'' effect of the neutrinos as they travel through the inhomogeneities of the gravitational 
potential. In the Boltzmann equation~(\ref{bolt-II}) it derives from Eq.~(\ref{II-pi}), which describes how 
the neutrinos momentum direction changes in time due to the potential wells they pass trough.  

Let us now continue our computation and apply the projection operator~(\ref{tensor-operator}) to 
Eq.~(\ref{en-momIIfourier}), keeping only the term~(\ref{only}) in $G_{\bf k}(\tau')$ and the ``pure'' tensor contribution 
$T_{\bf k}(\tau')$. We see that this
operator will act on the product of the two direction unit 
vectors $\hp^r \hp_s$ contained in $T^{r(\nu)}_{s}$.
Since $\delta^s_{~r} \hp^{ \, r} \hp_s=1$ and the product 
$\hat{\bf k} \cdot \hat{\bf p}$
defines the cosine $\mu$ between the direction along which the perturbation
propagates and the neutrino momentum, this means that 
$(\Pi^i_j)^{(\nu)}_{\bf k}$ will
 contain the term:
\be \label{tenspart}
\hp^i \hp_j - \mu (\hat{k}_j \hp^i +\hat{k}^i \hp_j ) 
+ \12 \hat{k}^i \hat{k}_j (1+\mu^2) - \12
 \delta^i_j (1- \mu^2).
\ee
 
As already done in Sec.~\ref{IIGWeq},  we can now choose one of the two 
polarization states
\mbox{$\lambda= + \, , \times$} and project the transverse 
traceless part we are interested in into 
the corresponding polarization tensors $\varepsilon^\lambda_{ij}$. Since a gravitational wave is transverse with respect to the
direction $\hat{{\bf k}}$ along which it propagates, we then have

\be
\label{enn}
\varepsilon_i^{\lambda j} \, \hp^i \hp_j= \frac{1}{\sqrt{2}}(1- \mu^2) (\delta^{+ \lambda} \cos2\varphi_{\bf n}+
\delta^{\times \lambda} \sin 2\varphi_{\bf n})\, ,
\ee
where the angle $\varphi_{\bf n}$ 
is the azimuthal angle of the neutrino momentum direction ${\bf n}$ in the orthonormal basis ${\bf e}$, 
${\bar {\bf e}}$ and ${\hat {\bf k}}$.  
In Fourier space the term~(\ref{only}) evaluated at $\tau'$ becomes (for simplicity we omit the convolution integral)
\begin{eqnarray}
& -& ({\bf k}_1 \cdot {\bf n}) ({\bf k}_2 \cdot {\bf n})(\Phi_{{\bf k}_2}(\tau')+\Psi_{{\bf k}_2}(\tau'))
\, q \dof \\
&\times&  \int_{\tau_{dec}}^{\tau'} d \tau''
e^{ik_1\mu_1(\tau''-\tau')} (\tau''-\tau') (\Phi'_{{\bf k}_1}(\tau'')+\Psi'_{{\bf k}_1}(\tau''))\, , \nonumber 
\end{eqnarray}
where we have written explicitly the term $\partial{F_{{\bf k}_1}} / \partial{\hp^i}$ 
as 
\begin{eqnarray}
\frac{\partial{F_{{\bf k}_1}}}{\partial{\hp^i}}&=& q \dof 
\int^{\tau}_{\tau_{dec}} d \tau'\,i k^i_1(\tau-\tau') [\fik'(\tau') +
 \psik'(\tau')] \nonumber \\
&\times& e^{i k_1 \mu_1 (\tau' -\tau )}\, ,
\end{eqnarray}
using the solution of Eq.~(\ref{fik1}). It proves convenient to take the tensor part of 
$({\bf k}_1 \cdot {\bf n}) ({\bf k}_2 \cdot {\bf n})$ which reads 
\begin{equation}
(k_{2l}k_{1m} \varepsilon^{+lm}) \varepsilon^{+}_{rs} n^r n^s+
 (k_{2l}k_{1m} \varepsilon^{\times lm}) \varepsilon^{\times}_{rs} n^r n^s\, .
\end{equation}   
Notice that this step is equivalent to follow the decomposition~(\ref{splitting}) of Ref.~\cite{kasaitomita} which allow to 
isolate the tensor contributions to the distribution function.\footnote{Also in this case we have verified that the scalar and 
vector components of $({\bf k}_1 \cdot {\bf n}) ({\bf k}_2 \cdot {\bf n})$ give a vanishing contribution 
once the energy momentum tensor is projected along the two polarization 
tensors~(\ref{tensorpolarizz}) and the angular integration in Eq.~(\ref{en-momIIfourier}) is performed.}

With these results we are now able to write the expression for 
$\Pi^{(\nu)}_{\bf k,\lambda}=
\varepsilon^{\lambda j}_{i}\Pi^{i(\nu)}_{j\bf k}$ at second order in 
the perturbations
\bea\label{Pik} 
\Pi^{(\nu)}_{{\bf k},\lambda}& = & - 2 a^{-4} g_i \int\!\!\!\int \frac{d^3 k_1 \, 
d^3 k_2}{(2 \pi)^3} \,
\delta({\bf k}_1 + {\bf k}_2-{\bf k}) \nonumber \\
&  \Bigg[ & \int \frac{d^3 q}{(2\pi)^3}\, q^2 \dof 
\int^{\tau}_{\tau_{dec}} {d \tau'}
  e^{ik\mu (\tau'-\tau)} \varepsilon_i^{\lambda j} \, n^i n_j \nonumber \\
&\times& [(k_{2l}k_{1m} \varepsilon^{+lm}) \varepsilon^{+}_{rs} n^r n^s+
 (k_{2l}k_{1m} \varepsilon^{\times lm}) \varepsilon^{\times}_{rs} n^r n^s] \nonumber \\
& \times & A_{{\bf k}_2}(\tau') \int_{\tau_{dec}}^{\tau'} d \tau''
e^{ik_1\mu_1(\tau''-\tau')} (\tau''-\tau') A'_{{\bf k}_1}(\tau'')\Bigg] \nonumber \\ 
&+&a^{-4} g_i \int \frac{d^3 q}{(2\pi)^3}\, \frac{q^2}{4}
\dof  (1- \mu^2)^2 \nonumber \int^{\tau}_{\tau_{dec}} d\tau' \nonumber \\ 
& & \chi'_{{\bf k},\lambda}(\tau')  e^{ik\mu (\tau'-\tau)} \, ,
\eea
where $A_{{\bf k}}(\tau) \equiv  \Phi_{{\bf k}}(\tau)+\Psi_{{\bf k}}(\tau)$. 

At this point we want to solve the angular and the momentum
dependence of the transverse, traceless neutrino
energy momentum tensor $\Pi^{(\nu)}_{\bf k}$. This implies solving the $d^3q$ integral in the previous expression. 

First of all notice that the pure tensor part in the
last line of Eq.~(\ref{Pik}) represents the second-order analogue of
the damping effect found and discussed in Refs.~\cite{Bond,RS,D,Weinb,Kamion,Repko,Komatsu,kasaitomita}. 
We are then able to deal with it (see, for example, Refs.~\cite{kasaitomita,Weinb,Komatsu}) and we find 
\be
 a^{-4}\int\frac{dq q^{4}}{(\pi)^{2}} 
\frac{\partial{\overline{F}}}{\partial{q}}
\int_{u_{dec}}^{u} dU \chi_{k}^{'} (U)
\frac{1}{15}[j_{0}(s)+\frac{10}{7}j_{2}(s)+\frac{3}{7} j_{4}(s)]\, ,
\ee
which becomes 
\be
- 8 \bar{\rho}_\nu(\tau) \int_{u_{dec}}^{u} dU \chi_{k}^{'} (U)
\frac{1}{15}[j_{0}(s)+\frac{10}{7}j_{2}(s)+\frac{3}{7} j_{4}(s)]\, ,
\ee
where we have used Eqs.~(\ref{unpdensity})-(\ref{fderivative}), 
the variable change $\tau \rightarrow u=k\tau$, $s=u-U$, with $U=k\tau'$, 
and the derivative is with respect to $U$. This means that also at second order gravitational waves
are damped by neutrino free streaming, as expected on general grounds. Here 
$\bar{\rho}_\nu$ is the unperturbed neutrino 
energy density given in Eq.~(\ref{unpdensity}).

We now have to deal with the reminder of Eq.~(\ref{Pik}), giving an 
additional source term that represents a completely new result. 

\subsubsection{Angular integration} \label{intang}
We notice that there is a $\mu$-dependence hidden in the
 exponential   
\begin{eqnarray}
\label{expexpansion}
e^{i \mu_1 s'}&=& 
\sum_{l=0}^{+\infty} i^l (2l+1)  j_l(s') P_l(\mu_1) \\
&=&
\sqrt{4 \pi} \sum_{l=0}^{+\infty} i^l j_l(s')\
\sum_{m=-l}^{+l} a_{l\,m} e^{im \varphi_{\bf n}}
  P_{lm}(\mu) Y^*_{lm}(\hat{\bf k}_1)\, , \nonumber 
\end{eqnarray}
where $a_{l\,m}=\sqrt{\frac{(2l+1)(l-m)!}{(l+m)!}}$ and 
$s'=k_1(\tau'-\tau) \equiv (U_1 - u_1)$.
For the integration over the angle $\varphi_{\bf n}$ we need to take into account 
the product $(\varepsilon_i^{\lambda j} \, \hp^i \hp_j) ( \varepsilon^{\lambda'}_{rs} n^r n^s)$ 
in Eq.~(\ref{Pik}) and, using Eq.~(\ref{enn}), we find 
\begin{eqnarray}
\label{intphi}
\int d\varphi_{\bf n} e^{i m \varphi_{\bf n}} 
(\varepsilon_i^{\lambda j} \, \hp^i \hp_j) ( \varepsilon^{\lambda'}_{rs} n^r n^s) = \frac{\pi}{2} (1-\mu^2)^2 
\delta^{\lambda \lambda'} \delta_{m0}\, , \nonumber \\
\end{eqnarray}
which means that the cross terms vanish, while the squared ones select $m=0$. 
When $m=0$ the $P_{lm}$ correspond to the Legendre
polynomials $P_l$ and we have:
\be
\sqrt{4 \pi} \sum_{l=0}^{+\infty} j_l(s') i^l \sqrt{2l+1}
Y^*_{l0}(\hat{\bf k}_1)P_{l}(\mu)  \, .
\ee
Therefore by expanding $e^{ik\mu(\tau'-\tau)}$ as in Eq.~(\ref{expexpansion}) 
we are led to an integration over $\mu$ of the 
following quantity
\begin{eqnarray}
\label{intmu}
\int_{-1}^{+1} d\mu \sum_{l l'} A_l P_l(\mu) B_{l'} P_{l'}(\mu) (1-\mu^2)^2\, , 
\end{eqnarray}
where
\begin{eqnarray}
\label{defAB}
A_l&=&i^l\sqrt{4 \pi(2l+1)} \, Y^*_{l0}(\hat{\bf k}_1) \, j_l[k_1(\tau''-\tau')]\, , \nonumber \\
B_{l'}&=&i^{l'} (2l'+1) \, j_{l'}[k(\tau'-\tau)]\, .
\end{eqnarray}
Such an integral is easily performed using the formulae of Appendix~\ref{Aformulae}, and Eq.~(\ref{intmu}) becomes
\begin{equation}
\label{fangular}
\sum_l \frac{1}{2l+1} \left[A_lB_l-A_lB^{(1)}_l-A^{(1)}_lB_l+A^{(1)}_lB^{(1)}_l \right]\, ,
\end{equation}
where 
\begin{eqnarray}
\label{defA1}
A^{(1)}_l&=&\frac{l(l-1)}{(2l-3)(2l-1)} A_{l-2} +\Bigg[ \frac{(l+1)^2}{(2l+1)(2l+3)} \nonumber \\
&+&\frac{l^2}{(2l+1)(2l-1)}\Big] A_l+
\frac{(l+2)(l+1)}{(2l+3)(2l+5)} A_{l+2}\, , \nonumber \\ 
\end{eqnarray}
and similar for $B^{(1)}_l$.

\subsubsection{Integration over the comoving momentum $q$.}

We can treat the momentum integration over $q$ in Eq.~(\ref{Pik}) 
independently from the angular
part. There is just one type of $q$-dependence, namely 
$q^4 \partial  \overline{F}/\partial q$. 
Remember that $\overline{F}$ is the unperturbed neutrino distribution
function given by the Fermi-Dirac distribution and 
\be
\label{unpdensity}
 \overline{\rho}_\nu =  a^{-4} \frac{g_i}{2 \pi^2 }\int \of q^3 dq
   = \frac{7}{8}   \frac{\pi^2}{30}T_{\nu}^4\, ,
      \ee
is the unperturbed neutrino energy density. Hence, integrating by parts when necessary, 
we find (the $\pi/2$ factor comes from Eq.~(\ref{intphi}))
\be
\label{fderivative}
a^{-4} g_i \frac{\pi}{2} \int \frac{dq}{(2 \pi)^3} \dof q^4  = -\frac{1}{2} \overline{\rho}_\nu \, .
\ee

\subsubsection{Final expression for the neutrino contribution}

Collecting the previous results we arrive at the contribution of 
streaming neutrinos to the evolution of gravitational waves:
\bea \label{new}
& & \Pi^{(\nu)}_{\bf k,\lambda} =  \overline{\rho}_\nu  
\int\!\!\!\int \frac{d^3 k_1 \, d^3 k_2}{(2 \pi)^3}  \delta({\bf k}_1 + {\bf k}_2-{\bf k}) (k_{2r}k_{1s}\varepsilon^{rs}_\lambda) 
\nonumber \\
& & \sum_l  \frac{1}{2l+1} \int_{\tau_{dec}}^\tau d\tau' (\Phi_{{\bf k}_2}(\tau')+\Psi_{{\bf k}_2}(\tau')) \left[
B_l-B^{(1)}_l\right] \nonumber \\
& & \times \int_{\tau_{dec}}^{\tau'} d\tau''(\tau''-\tau') (\Phi'_{{\bf k}_1}(\tau'')+\Psi'_{{\bf k}_1}(\tau''))
\left[A_l-A^{(1)}_l\right] \nonumber \\
& &-  8  \bar{\rho}_\nu(\tau) \int_{\tau_{dec}}^{\tau} d \tau' \chi_{k}^{'} (\tau')
\frac{1}{15}[j_{0}(s)+\frac{10}{7}j_{2}(s)+\frac{3}{7} j_{4}(s)] \nonumber \\
\eea
where $s=k(\tau-\tau')$, and the functions $A_l$, $A^{(1)}_l$, $B_l$ and $B^{(1)}_l$ are defined 
in Eq.~(\ref{defAB}) and Eq.~(\ref{defA1}).

This equation is clearly telling us that collisionless free
streaming neutrinos contribute with  new terms to the source of 
second-order gravitational waves, with respect to the 
fluid treatment adopted in the literature so far, where the tensor 
part of $\Pi^{i(\nu)}_{~j}$ at second order has 
never been taken into account.  

Even with a
qualitative approach, we can state that if low $l$
contributions (up to $l=2$) can have a correspondence with respect to
a source where neutrinos and photons are treated as a single-fluid radiation,
higher multipoles surely do not. These in fact come from the high
neutrino velocity dispersion 
and can be found only if neutrinos are treated as
collisionless particles. 

Moreover, during the radiation-dominated epoch, the neutrino fraction
$f_\nu(\tau)=\Omega_\nu/\Omega_R$ 
is not negligible. Since all the terms in $\Pi^{(\nu)}_{\bf k}$ are multiplied
by $\overline{\rho}_\nu$, they are therefore non-negligible at that time.\\

%%%%%%%%%%%%%%%%%%%%%%%%%%%%%%%%%%%%%%%%%%%%%%%%%%%%%%%%%%%%%%%%%%%%%%%%%%%%%%
\section{The photon tensor quadrupole}
\label{photons+neutrinos}
%%%%%%%%%%%%%%%%%%%%%%%%%%%%%%%%%%%%%%%%%%%%%%%%%%%%%%%%%%%%%%%%%%%%%%%%%%%%

In order to complete the expression of the source term for
gravitational waves during the
radiation era in Eq.~(\ref{eqevol}), we now have 
to add the photon contribution.
Following \cite{cmbanisotr}, the second-order photon
quadrupole in the tight coupling limit is given by:
\be
\Pi^{(2)ij}_\gamma \simeq \frac{8}{3} \left(v^{(1)i}_\gamma
v^{(1)j}_\gamma - \frac{1}{3}
\delta^{ij} v^{(1)2}_\gamma  \right)\,.
\ee

To get the photon contribution to the second-order gravitational 
waves source we then have to extract the 
transverse and traceless component.

Since the operator to apply is such that 
${\cal P}_{rj}^{~is} \delta^r_s=0$, we will
find that the tensor quadrupole takes the form:
\be
\Big( \Pi^{(2)i}_{\gamma ~j} \Big)_{TT} \simeq \frac{8}{3} 
{\cal P}_{rj}^{~is} v^{(1) r}_\gamma v^{(1)}_{\gamma \, s}\, .
\ee

By making use of the first-order space-time component of the Einstein
equations (see Appendix~\ref{1orderEinstein})
\be
\frac{1}{a^2} \partial^i ( \Psi' + \mathcal H \Phi)=- \frac{ 16 \pi G}{3}
(\overline{\rho}_\gamma  v^{(1) i}_\gamma + 
\overline{\rho}_\nu  v^{(1) i}_\nu) \, ,
\ee
we can define the first-order photon velocity as:
\be
 v^{(1) i}_\gamma= - \Big[ \frac{3}{16 \pi G a^2 \overline{\rho}_\gamma}
 \partial^i ( \Psi' + \mathcal H
   \Phi) + \frac{\overline{\rho}_\nu}{\overline{\rho}_\gamma} v^{(1) i}_\nu
\Big] \, ,
\ee
where $ v^{(1) i}_\nu$ represents the first-order neutrino velocity.
During the radiation-dominated epoch neutrinos are still relativistic
so that, in Fourier space, the velocity can be written as:
\be
 v^{(1) i}_\nu= \frac{1}{\rho_\nu+ P_\nu} \int \frac{d^3 p}{(2 \pi)^3}
 F_\nu^{(1)} p^i = -12 \pi i \mathcal{N}_1 \hat{k}^i.
\ee
The term $\mathcal{N}_1$ refers to the neutrino dipole and $P_\nu$ is
the neutrino pressure.
By making use of the formal solution (\ref{fik1}) of the first-order neutrino
Boltzmann equation, the dipole can be expressed explicitly in
terms of the perturbations to give:
\bea
\mathcal{N}_1(k)&=& \frac{i}{2} \frac{\partial{\ln
\overline{F}_\nu}}{\partial{\ln q}} \int^\tau_{\tau_{dec}}d \tau' 
               \Big[\frac{i k_1}{3} \fik'(j_0 (s) 
                    - 2 j_2 (s)) \nonumber \\
&+& i \psik' j_1 (s)
               \Big]\, .
\eea

Projecting along one of the polarization state ($+$), we have that, in
Fourier space, the photon contribution we are searching for takes the
form:
\bea\label{photonquadrupole}
\Pi^{(\gamma) \lambda}_{\bf k} &=& \int 
\frac{d^3 k_1 d^3 k_2}{(2 \pi)^3}\delta({\bf k}_1+{\bf k}_2-{\bf k}) 
 \varepsilon^{j \, \lambda}_{i} k_{1 \, j} k^i_2 \nonumber\\
        &\times&
       \Big[\frac{3 k_1 k_2}{32 (\pi G a^2 \overline{\rho}_\gamma)^2}
	  (\psik' + \mathcal H \fik) (\psikd' + \mathcal H \fikd)
	  \nonumber \\
          &+&32 \cdot 12 \left(\frac{\overline{\rho}_\nu
	    \pi}{\overline{\rho}_\gamma}\right)^2
            \mathcal{N}_1(k_1)\mathcal{N}_1(k_2) \\
          &+& 6 \pi \frac{\overline{\rho}_\nu}{
	      \overline{\rho}_\gamma} k_2 \mathcal{N}_1(k_1)(\psikd'
	    + \mathcal H \fikd)\nonumber\\
          &+&  \frac{6 \overline{\rho}_\nu}{G a^2
	      \overline{\rho}^2_\gamma} k_1 \mathcal{N}_1(k_2)(\psik'
	    + \mathcal H \fik)\nonumber 
       \Big] \, .
\eea 
It is worth stressing that this expression 
contains at most multipoles with $l=2$. Therefore, if we had used a fluid
treatment for the neutrinos as well (e.g. $\Pi^{i j(2)}_\nu \propto v^{(1)
  i}_\nu v^{(1)j}_{\nu} $), we would have found in the
second-order neutrino contributions terms with $l$ not higher then
$l=2$. This is clearly in contrast with the 
neutrino source term Eq.~(\ref{new}).

\section{Conclusions}
\label{Concl}

This paper represents the first step towards the quantitative evaluation of 
the impact of cosmic neutrinos on the evolution of the  
gravitational wave background; it provides, first of all, a complete study 
of the Boltzmann equations for neutrinos at second order and the 
expression for the second-order anisotropic stress tensor.  

Free-streaming neutrinos are an important source of second-order 
gravitational waves during the radiation-dominated epoch. Along with the fact 
that neutrinos yield a relevant contribution to the total energy 
during this epoch, this is due to the large neutrino velocity
dispersion and it emerges from the calculations performed in this paper 
when assuming that neutrinos are collisionless particles, as it is 
the case after their decoupling. 

The fluid treatment adopted so far for describing neutrinos turns out 
therefore to be a poor approximation that leads, in particular, 
to underestimating the role of neutrino free-streaming as a source of 
gravitational waves. In particular, in this 
paper we have made the first full consistent computation of the second-order 
tensor part of the neutrino 
anisotropic stress tensor, Eq.~(\ref{new}). 
This has been achieved by computing and solving the second-order Boltzmann 
equation for the neutrino  distribution function. 
Besides recovering the second-order counterpart of the damping effect 
studied in Ref.~\cite{Bond,RS,D,Weinb,Kamion,Repko,Komatsu,kasaitomita}, Eq.~(\ref{new}) 
represents a completely new source term for the evolution of 
gravitational waves.    

% As a second step of our analysis,  
% a numerical calculation is needed to precisely assess the relevance of this
% effect \cite{inpreparation}. In doing this it is necessary to 
% truncate at some $\ell_{max}$ the infinite multipoles coming from the expansion of
% the first-order Boltzmann equation into an infinite hierarchy of moment
% equations. This is similar to what happens in CMB studies, where the value of 
% $\ell_{max}$ increases as the mass of the neutrinos decreases (see for example Ref.~\cite{Bert}).

%Should we take as a reference whaone does in CMB studies, using the 
%truncation schemes described in Ref.~\cite{Bert} one could expect  
%$l_{max} \sim 100$ to be roughly adequate at better than the $0.1\%$
%level for massive neutrinos with mass in the
%range of 1 eV. It is worth noticing that in the case of massless
%neutrinos we would have found $\ell_{max} \sim 1000$. Higher multipole
%moments then decay rapidly once the neutrinos become non-relativistic.

\section*{Acknowledgments}
We thank C. Carbone for useful discussions.
This research has been partially supported by ASI contract I/016/07/0
"COFIS" and ASI contract Planck LFI Activity of Phase E2.
This research was also supported in part by
the Department of Energy and the European Community's Research Training
Networks under contracts MRTN-CT-2004-503369, MRTN-CT-2006-035505.

%%%%%%%%%%%%%%%%%%%%%%%%%%%%%%%%%%%%%%%%%%%%
%%%%%%%%%%%%%%%%%%%%%%%%%%%%%%%%%%%%%%%%%%%

\addcontentsline{toc}{chapter}{Appendix}

\appendix
\section{The second-order Einstein tensor} 
\label{christoffel}

In this appendix we provides the definitions for the connection coefficients and 
the expression of the second-order Einstein tensor for the metric~(\ref{poisson}): 
\be
ds^2=a^2(\tau)\left[-e^{2\Phi} d\tau^2+2\omega_i dx^i d\tau +(e^{-2\Psi}\delta_{ij}\right. + \left. \chi_{ij}) dx^i dx^j\right].\nonumber
\ee 

The space-time metric $g_{\mu \nu}$ has signature ($-,+,+,+$). 
The connection coefficients are defined as
\begin{equation}
\label{conness} \Gamma^\alpha_{\beta\gamma}\,=\,
\frac{1}{2}\,g^{\alpha\rho}\left( \frac{\partial
g_{\rho\gamma}}{\partial x^{\beta}} \,+\, \frac{\partial
g_{\beta\rho}}{\partial x^{\gamma}} \,-\, \frac{\partial
g_{\beta\gamma}}{\partial x^{\rho}}\right)\, .
\end{equation}
Greek indices ($\alpha, \beta, ..., \mu, \nu, ....$)
 run from 0 to 3, while latin indices ($a,b,\dots,i,j,k,\dots$,
$m,n,\dots$) run from 1 to 3. 
In particular, their explicit expression with our metric reads:
\noindent 
\begin{eqnarray}
\Gamma^0_{00}&=& {\mathcal H}+\Phi'\, ,\nonumber\\
\Gamma^0_{0i}&=& \frac{\partial\Phi}{\partial x^i}+
{\mathcal H}\omega_i\, ,\\
\Gamma^i_{00}&=& \omega^{i'}+{\mathcal H}\omega^i+e^{2\Psi+2\Phi}
\frac{\partial\Phi}{\partial x_i}
\, ,\nonumber\\
\Gamma^0_{ij}&=& -\frac{1}{2}\left(\frac{\partial \omega_j}{\partial x^i}+
\frac{\partial \omega_i}{\partial x^j}\right)+e^{-2\Psi-2\Phi}
\left({\mathcal H}-\Psi'\right)\delta_{ij}\nonumber\\
&+&\frac{1}{2}\chi_{ij}'+
{\mathcal H}\chi_{ij}\, ,\nonumber\\
\Gamma^i_{0j}&=&\left({\mathcal H}-\Psi'\right)\delta_{ij}+
\frac{1}{2}\chi_{ij}'+\frac{1}{2}\left(\frac{\partial \omega_i}{\partial x^j}-
\frac{\partial \omega_j}{\partial x^i}\right)\, ,\nonumber\\
\Gamma^i_{jk}&=&-{\cal H}\omega^i\delta_{jk}-
\frac{\partial\Psi}{\partial x^k}\delta^i_{~j}-
\frac{\partial\Psi}{\partial x^j}\delta^i_{~k}+
\frac{\partial\Psi}{\partial x_i}\delta_{jk} \nonumber \\
& + & 
\frac{1}{2} \left(\frac{\partial\chi^i_{~j}}{\partial x^k}+
\frac{\partial\chi^i_{~k}}{\partial x^j} - 
\frac{\partial\chi_{jk}}{\partial x_i} \right) \, \nonumber .
\end{eqnarray}

The Einstein equations are written as 
$G_{\mu\nu}= 8\pi G_{\rm N} T_{\mu\nu}$,
where $G_{\rm N}$ is the usual Newtonian gravitational constant, $G_{\mu\nu}=R_{\mu\nu}-\frac{1}{2}g_{\mu\nu}R$ is the Einstein tensor and $T_{\mu\nu}$ is the Energy-momentum tensor.

The Ricci tensor $R_{\mu\nu}$ is a contraction of the Riemann tensor, 
$R_{\mu\nu}=R^{\alpha}_{~\mu\alpha\nu}$ 
and in terms of the connection coefficient it is given by
\begin{equation}
R_{\mu\nu}\,=\, \partial_\alpha\,\Gamma^\alpha_{\mu\nu} \,-\,
\partial_{\mu}\,\Gamma^\alpha_{\nu\alpha} \,+\,
\Gamma^\alpha_{\sigma\alpha}\,\Gamma^\sigma_{\mu\nu} \,-\,
\Gamma^\alpha_{\sigma\nu} \,\Gamma^\sigma_{\mu\alpha}\,.
\end{equation}
The Ricci scalar is the trace of the Ricci tensor, 
$R=R^{\mu}_{~\mu}$. 
%%%%
\noindent
The components of Einstein's tensor up to second-order read 
\begin{eqnarray}
\label{LH200}
G^0_{~0}&=&-\frac{e^{-2\Phi}}{a^2}\left[3{\mathcal H}^2-6{\mathcal H}\Psi'
+3(\Psi')^2 \right. \nonumber \\
&-& \left. e^{2\Phi+2\Psi}\left(\partial_i\Psi\partial^i\Psi-2\nabla^2
\Psi\right)\right]\, ,\\
\label{LH2i0}
G^i_{~0}&=&2\frac{e^{2\Psi}}{a^2}\left[\partial^i\Psi'+\left({\mathcal H}-
\Psi'\right)\partial^i\Phi\right]-\frac{1}{2a^2}\nabla^2\omega^i \nonumber 
\\ 
&+& \left(4{\mathcal H}^2-2\frac{a''}{a}\right)\frac{\omega^i}{a^2}\, ,\\
\label{LH2ij}
G^i_{~j}&=&\frac{1}{a^2}\biggl[e^{-2\Phi}\biggl({\mathcal H}^2-2\frac{a''}{a}-
2\Psi'\Phi'-3(\Psi')^2+\\
&+& 2{\mathcal H}\bigl(\Phi'+2\Psi'\bigr)+2\Psi''\biggr) 
+ e^{2\Psi}\biggl(\partial_k\Phi\partial^k\Phi+\nonumber\\
&+& \nabla^2\Phi-\nabla^2\Psi
\biggr)\biggr]\delta^i_j 
+ \frac{e^{2\Psi}}{a^2} 
\biggl(-\partial^i\Phi\partial_j\Phi  -
\partial^i\partial_j\Phi+\nonumber\\
&+&\partial^i\partial_j\Psi- 
\partial^i\Phi\partial_j\Psi
+\partial^i\Psi\partial_j\Psi-\partial^i\Psi\partial_j\Phi\biggr)\nonumber\\
&-&\frac{{\mathcal H}}{a^2}\left(\partial^i\omega_j+\partial_j\omega^i\right)
-\frac{1}{2a^2}\left(\partial^{i}\omega_j'+\partial_j\omega^{i'}\right)
\nonumber \\
&+& \frac{1}{a^2}\left(
{\mathcal H}\chi^{i'}_j+\frac{1}{2}\chi_j^{i''}-\frac{1}{2}\nabla^2
\chi^i_j\right) \, ,\nonumber
\end{eqnarray}
The exponentials are maintained since they help in simplifying lots of calculation, however notice that in all these expressions they are implicitly truncated up to second order.

\section{First-order perturbations of Einstein equations for 
photons and neutrinos}
\label{1orderEinstein}

Carrying out the calculation, the first-order Einstein's 
equations, expressed in Fourier space in terms of 
the perturbations $\Phi$ and $\Psi$, take the form:
\be
-k^2 \Psi - 3 \frac{\dot a}{a} \left(\dot \Psi - \Phi \frac{\dot a}{a}
\right)= 16 \pi G a^2 [\overline{\rho}_\gamma  
\Theta_{\gamma 0} +\overline{\rho}_\nu \Theta_{\nu 0} ]
\ee
\be
k^2(\Phi - \Psi)= - 32 \pi G a^2 \overline{\rho}_\nu \Theta_{\nu 2}.
\ee

Here $\Theta_{\gamma 0}$ and $\Theta_{\nu 0}$ are, respectively, the photon
and the neutrino monopole contribution, while $\Theta_{\nu 2}$ refers to the
neutrino scalar quadrupole.

Remember that the first-order perturbation to the neutrino distribution 
function is defined by 
$F^{(1)}_{\nu}= \overline{F}_\nu \mathcal N$,
where
\be
\mathcal N= - 
\frac{\partial{ \ln \overline F_{\nu}}}{\partial{\ln p}}\Theta_\nu
\ee
and $\Theta_\nu(t, \vec x, \hp)=
\frac{\delta T_\nu}{T_\nu}$. Similarly it happens
to the photon contribution.

In general, the $l$th multipole of the
temperature field $\Theta$ can be defined as:
\be
\Theta_{l}= \frac{1}{(-i)^l} \int^1_{-1} \frac{d \mu}{2}
\mathcal{P}_l(\mu) \Theta\, ,
\ee 
where $\mathcal{P}_l(\mu)$ are the Legendre polynomials.

The neutrino monopole is then defined by:
\be
\Theta_{\nu \,0}= \int^1_{-1} \frac{d \mu}{2} \Theta_\nu\, ,
\ee
while the quadrupole, corresponding to $l=2$, is
\be
\Theta_{\nu \,2}= -  \int^1_{-1} \frac{d \mu}{2}
\mathcal{P}_2(\mu) \Theta_\nu\, .
\ee
 
\section{The second-order neutrino Boltzmann equation}
\label{appendixBoltzmannII}

In this appendix we provide the explicit form for the different contributions in the neutrino Boltzmann equation 
 $ \frac{\partial{F_{\nu}}}{\partial{\tau}}
+\frac{\partial{F_{\nu}}}{\partial{x^{i}}}\cdot\frac{dx^{i}}{d \tau}
+\frac{\partial{F_{\nu}}}{\partial{E}}\cdot\frac{dE}{d \tau} 
+\frac{\partial{F_{\nu}}}{\partial{n^i}}\cdot\frac{d n^i}{d\tau}
=0 \,$:

\begin{itemize}

\item $ \frac{dx^{i}}{d \tau} =  \frac{dx^{i}}{d \lambda} \, \frac{d \lambda}{d \tau} \equiv  \frac{P^i}{P^0} $.

For this term we have:
\be \label{dxi}
\frac{dx^{i}}{d \tau} = \, \pE \, \hp^i e^{(\Psi+\Phi)} [1- \omega_j
  \, \hp^j \,
  \pE \ - \12 \chiij]
\ee

\item $\frac{d E}{d \tau}$.

Deriving an expression for this term is more lengthy since it involves
the use of the geodesic equation. 

We can start noting that, using Eq.~(\ref{4mom2}), we obtain:
\be
\frac{d P^0}{d \tau}= e^{- \Phi} \, \frac{E}{a}
\Big[- \frac{d \Phi}{d \tau} B + \frac{d B}{d \tau} 
+ B \Big(\frac{1}{E}\frac{d E}{d \tau}
 - \frac{1}{a}\frac{d a}{d
    \tau} \Big) \Big] \;,
\ee
where $B= 1 + \pE \omega_i \hp^i$.

To simplify a bit the notation, from now on we will set
\be
g' \equiv \frac{\partial g}{\partial \tau} \quad \textrm{and} \quad
g_{,j} \equiv \frac{\partial g}{\partial x^j},\quad \forall \, g
\equiv g(\tau, x^i)\, .
\ee

If we now make the total derivatives explicit, according to the fact that
\be
\frac{d \Phi}{d \tau}= \Phi' + \Phi_{, \, i} \frac{P^i}{P^0}\, ,
\ee
and
\be
\frac{d B}{d \tau} \equiv \pE \, \hp^i \frac{d \omega_i}{d \tau} =\,
\pE \, \hp^i \Big(\omega'_i + \omega_{i,j}\frac{P^j}{P^0} \Big)\, ,
\ee
we find:
\bea
\frac{1}{E}\frac{d P^0}{d \tau}&=& \frac{e^{- \Phi}}{a} (1+ \pE \,
\omega_i \hp^i) \Big[ - (\Phi' + \Phi_{,i} \frac{P^i}{P^0})
  \nonumber\\
&+& \frac{d E}{d \tau} - 
      \frac{1}{a} \frac{d a}{d \tau}+\pE \, \hp^i (\omega'_i +
      \omega_{i,j}\frac{P^j}{P^0})\Big] \, .
\eea
By making use of the geodesic equation, we can express
the time component as a sum of three terms
\bea
& &   -\Gamma^{0}_{\alpha\beta}\frac{P^{\alpha}P^{\beta}}{P^0} =
 -\Gamma^{0}_{00}P^{0}
 - 2 \Gamma^0_{0 j}P^{j}
-\Gamma^{0}_{i j} \frac{P^{\alpha}P^{\beta}}{P^0}\, . \nonumber \\
\eea
If we now insert the connection coefficients 
and the perturbed components of the 4-momentum (\ref{4mom1}) and (\ref{4mom2}), we come
at last to the required term
\bea\label{dEtot}
\frac{1}{E}\frac{d E}{d \tau}&=& -\, \Big(\pE \Big)^2 \mathcal H - \pE \hp^j
\Phi_{,j} e^{ \phi + \Psi}
+ \Big(\pE \Big)^2 \Psi' \nonumber \\
&-& \, \pE \Big[\hp^i \omega'_i+ 2 \hp^j \omega_j \mathcal H 
- 2 \mathcal H
  \omega_r \hp^r \left(\pE\right)^2   \Big] 
\nonumber \\
&-&  \Big(\pE \Big)^2 \12 \chiijp \, .
\eea

We can note that the first term in the RHS of 
the latter equation is the zeroth-order time component of the geodesic
equation. 

\item $\frac{d n^i}{d \tau}$.

This term requires some lengthy algebra as well. To obtain it the
spatial component of the geodetic equation must be used:
\be
\frac{d P^i}{d \tau} =  
-\Gamma^{i}_{\alpha\beta}\frac{P^{\alpha}P^{\beta}}{P^0}.
\ee

Carrying on the calculation in the same way as we did for 
$d E/d \tau$, at the end we recover the expression
\bea
\frac{d P^i}{d \tau} &=&
 \frac{p}{a} \Big[ \frac{d \hp^i}{d \tau} +
  \hp^i \frac{d \Psi}{d \tau} 
+ \frac{\hp^i}{p} e^\Psi  \Big(-2 \pE E
  \mathcal H - \hp^j \Phi_{,j} E \nonumber \\
&+& \pE E \Psi'  \Big) \Big]\, .
\eea

The term we are looking for has the form
\be\label{II-pi}
\frac{d \hp^i}{d \tau}= -\frac{E}{p} 
\Phi^{,i} - \frac{E}{p} \Psi^{,i} 
                        + \hp^i \hp^j \frac{E}{p} \Phi_{,\, j} + \pE
			\hp^i \hp^k \Psi_{,\, k}\, .
\ee
It is worth noticing that, since
$\partial{\fnu}/\partial{\hp^i}$ is already a first-order term, we
must consider this equation just up to first-order. The scalars that appear
here are therefore the linear components of the perturbations.

\end{itemize} 
The purpose of this Appendix is to find the second-order Boltzmann
equation for (decoupled) neutrinos. It is then useful to use the relations in
Eqs.~(\ref{II-distrib}) and (\ref{II-fipsi}) for making the second-order
terms explicit.

The second-order contributions of each part of equation (\ref{bolt-II})
are listed below in the following expressions:

\be
\dfxi \frac{d x^i}{d \tau} \bigg\arrowvert_{2^{nd}\,\textrm{ord.}}
=\frac{\partial \fI}{\partial x^i}\, \pE
\hp^i (\fiI + \psiI) + \12 \frac{\partial \fII}{\partial x^i} \, \pE \hp^i\, .
\ee

The energy dependence term is a bit more complicated since it consists
of three terms such that:

\bea
\dfe \frac{d E}{d \tau} \bigg\arrowvert_{2^{nd}}
&=& \dofe \left( \frac{d E}{d \tau} \right) \bigg\arrowvert_{2^{nd}}
+\frac{\partial \fI}{\partial E}  \left(\frac{d E}{d \tau}\right)
\bigg\arrowvert_{1^{st}} \nonumber\\
&+& \12 \frac{\partial \fII}{\partial E} 
\left(\frac{dE}{d\tau}\right) \bigg\arrowvert_{0^{th}} \, .
\eea
(The overline refers to the zeroth-order neutrino distribution
function).

We then have

\bea
\frac{1}{E} \left( \frac{d E}{d \tau} \right)\bigg\arrowvert_{2^{nd}}&=&
-\pE \hp^j \left[ \12 \fiII_{,\, j} + \fiI_{,\, j} \psiI 
+ \fiI_{,\, j}
  \fiI \right]\nonumber\\
&+&\left( \frac{p}{E} \right)^2 \12 \psiIIp- \, V_{II}\, 
-\left( \frac{p}{E} \right)^2 \12 \chiijp \, , \nonumber \\
\eea

while

\be \label{dEI}
\frac{1}{E} \frac{d E}{d \tau} \bigg\arrowvert_{1^{st}}
= - \pE \hp^j \fiI_{,\, j} + \left( \frac{p}{E} \right)^2 \psiIp\, ,
\ee
 
and

\be
\frac{1}{E} \frac{d E}{d \tau} \bigg\arrowvert_{0^{th}}=
- \left( \frac{p}{E} \right)^2 \mathcal H\, .
\ee

The term $V_{II}$ accounts for the second-order vector
contribution. This already appears in Eq.~(\ref{dEtot}) and is defined as 
\bea
\label{VII}
V_{II}&=& \pE \Big[\hp^i \omega'_i+ 2 \hp^j \omega_j \mathcal H 
- 2 \mathcal H
  \omega_r  \hp^r \left( \pE \right)^2  \Big]\, .
\eea
 
Finally, we use Eq.~(\ref{II-pi}) to deal with the dependence on the
momentum direction. Matching together all these terms in Eq.~(\ref{bolt-II}), 
we have now all the tools needed to obtain the second-order Boltzmann 
equation.

\section{Some formulae used for the angular integration}
\label{Aformulae}

The coefficients $A^{(1)}_l$ and $B^{(1)}_l$ appearing in Eq.~(\ref{fangular}) are simply given in terms of Eq.~(\ref{defAB})
as 
\begin{eqnarray}
A^{(1)}_l&=&\frac{l(l-1)}{(2l-3)(2l-1)} A_{l-2} +\Bigg[ \frac{(l+1)^2}{(2l+1)(2l+3)} \nonumber \\
&+&\frac{l^2}{(2l+1)(2l-1)}\Big] A_l+
\frac{(l+2)(l+1)}{(2l+3)(2l+5)} A_{l+2}\, , \nonumber \\ 
\end{eqnarray}
and similar for $B^{(1)}_l$. 
The result in Eq.~(\ref{fangular}) is obtained using the orthogonality of the Legendre Polynomials and applying to 
Eq.~(\ref{intmu}) the formula (here $a_l$ is a generic function of $l$)
\begin{eqnarray}
\sum_l a_l j_l(x) \mu^2 P_l(\mu)=\sum_l {\widetilde a}_l P_l(\mu)\, ,
\end{eqnarray}
where 
\begin{eqnarray}
& & \sum_l {\widetilde a}_l=
\frac{l(l-1)}{(2l-3)(2l-1)} a_{l-2} j_{l-2}(x) +\Bigg[ \frac{(l+1)^2}{(2l+1)(2l+3)}\nonumber \\
& &+ \frac{l^2}{(2l+1)(2l-1)}a_l j_l(x)\Bigg]+
\frac{(l+2)(l+1)}{(2l+3)(2l+5)} a_{l+2} j_{l+2}(x)\, , \nonumber \\
\end{eqnarray}
which derives from the recursion relation of the Legendre Polynomials (see also~\cite{Tomitanew}).

\end{document}